\documentstyle[12pt,epsfig]{article}
\begin{document}

\begin{titlepage}
\rightline{July 2013}
\vskip 2cm
\centerline{\Large \bf
Tully-Fisher relation, galactic rotation curves 
}
\vskip 0.4cm
\centerline{\Large \bf
and dissipative mirror dark matter
}

\vskip 1.7cm
\centerline{R. Foot\footnote{
E-mail address: rfoot@unimelb.edu.au}}

\vskip 0.7cm
\centerline{\it ARC Centre of Excellence for Particle Physics at the Terascale,}
\centerline{\it School of Physics, University of Melbourne,}
\centerline{\it Victoria 3010 Australia}
\vskip 3cm
\noindent
If dark matter is dissipative then the distribution
of dark matter within galactic halos can be governed by dissipation, heating and 
hydrostatic equilibrium.
Previous work has shown that a specific model, in the framework of mirror dark matter, 
can explain several empirical galactic scaling relations. 
It is shown here that this  dynamical halo model 
implies a quasi-isothermal dark matter density,
$\rho (r) \simeq \rho_0 r_0^2/(r^2 + r_0^2)$, where
the core radius, $r_0$, scales with disk scale length, $r_D$, via
$r_0/{\rm kpc} \approx 1.4\left(r_D/{\rm kpc}\right)$. 
Additionally,  the product $\rho_0 r_0$ is roughly
$constant$, i.e.
independent of galaxy size
(the $constant$ is set by the parameters of the model).
The derived dark matter density profile implies that the galactic rotation velocity satisfies the Tully-Fisher relation,
$L_B \propto v^{3}_{max}$, where $v_{max}$ is the maximal rotational velocity.
Examples of rotation curves resulting from this dynamics are given.

\end{titlepage}


Small scale structure, that is, the structure of dark matter galaxy halos
has been a long standing puzzle.
If dark matter is
dissipative then there is 
a possible solution. Dissipative dark matter leads to nontrivial dynamics:
the halo of a spiral galaxy can then be governed by hydrostatic equilibrium,
dissipation and heating \cite{sph}.
Mirror dark matter (see \cite{review,review2} for reviews) offers a specific framework in which to study this possibility.
Within the mirror dark matter context a previous study \cite{foot5} has shown that
this emerging picture of halo dynamics can
potentially explain small scale structure 
as it leads to successful galactic scaling relations.
The purpose of this article is to further explore this dynamics.

Mirror dark matter supposes the existence of a hidden sector exactly isomorphic to the standard model \cite{flv}. 
That is, the Lagrangian governing the fundamental properties of the elementary particles has the form:
\begin{eqnarray}
{\cal L} = {\cal L}_{SM} (e, \nu, u, d, \gamma, ...) + {\cal L'}_{SM} (e', \nu', u', d', \gamma',...) + {\cal L}_{mix} \ .
\end{eqnarray}
Such a theory features an exact $Z_2$ symmetry, which can be interpreted as the spacetime parity symmetry, 
which maps each ordinary particle onto a mirror partner, denoted with a prime ($'$). 
This $Z_2$ symmetry is assumed to be unbroken. It follows that the mirror particles
each have the same mass and interactions among themselves as the particles in the ordinary sector.
The mirror quarks form mirror baryons which together with mirror electrons constitutes the inferred dark matter in the Universe \cite{review,review2}.
In addition to gravity, the mirror particles can interact with the ordinary particles
via kinetic mixing interaction:
\begin{eqnarray}
{\cal L}_{mix} = \frac{\epsilon}{2} F^{\mu \nu} F'_{\mu \nu}
\label{kine}
\end{eqnarray}
where $F_{\mu \nu}$ ($F'_{\mu \nu}$) is the field strength tensor for the photon (mirror photon).
Such kinetic mixing is gauge invariant and renormalizable and $\epsilon$ can be viewed as a fundamental
parameter of the theory \cite{he}.
The physical effect of the kinetic mixing interaction is to induce a tiny ordinary
electric charge ($\propto \epsilon$) for the hidden sector $U(1)'$ charged particles \cite{holdom}.

In this picture,
galactic halos are composed, predominantly, of a plasma of such self interacting 
and dissipative dark matter \cite{sph,foot5}.
If this is indeed the case, then
the halo could be modelled as a fluid, governed by Euler's equations of fluid dynamics. 
At the current epoch, this fluid is presumed to have evolved to a hydrostatic equilibrium configuration where
the energy being absorbed in each volume element is equal to the energy being
radiated from the same volume element.
Thus, in addition to the hydrostatic equilibrium condition we have the dynamical condition:
\begin{eqnarray}
{d^2 E_{in} \over dt dV}= 
{d^2 E_{out} \over dt dV} 
\ .
\label{meal8}
\end{eqnarray}
In ref.\cite{foot5}  approximate formulas for the left and right-hand sides 
of the above equation have been derived, which we summarize below.

\vskip 1.2cm
\noindent
{\it (a) dissipation}
\vskip 0.3cm
\noindent

We assume that thermal bremsstrahlung is the main dissipative process.
The rate at which bremsstrahlung energy is radiated per unit volume, per unit time 
at a particular point, $P$, 
in the halo
is \cite{bookastro}:
\begin{eqnarray}
{d^2 W \over dt dV} = {16\alpha^3 \over 3m_e}\left({
2\pi T \over 3m_e}\right)^{1/2} 
\ \sum_j
\left[ Z_j^2 n_j n_{e'} \bar g_B  \right]
\label{john}
\end{eqnarray}
where the index $j$ runs 
over the mirror ions in the plasma (of charge $Z_j$)  and
$\bar g_B$ is the frequency average of the velocity averaged
Gaunt factor for free-free emission. 
We take $\bar g_B = 1.2$, which is known to be 
accurate to within about 20\% \cite{bookastro} .
In our numerical work we approximate $E_{out} = W$. See ref.\cite{foot5} for further discussions.

\vskip 0.3cm
\noindent
{\it (b) heating}
\vskip 0.3cm

If kinetic mixing is nonzero (i.e. $\epsilon \neq 0$) then
light mirror particles, $e', \bar e', \gamma'$ can be produced in a 
core-collapse supernova from processes such as $e \bar e \to e' \bar e'$. 
In fact, mirror particle emission can be 
comparable to that of neutrinos for $\epsilon \sim 10^{-9}$ \cite{raffelt,sil}.
The bulk of this energy is expected to be carried off by mirror photons, $\gamma'$, in the region around 
an ordinary supernova.
The idea is that these mirror photons, with total energy up to around half the supernova core-collapse
energy ($\sim 10^{53} \ {\rm erg}$ per supernova) will replace the energy lost in the halo due to dissipation.
This heating is achieved by interactions (photoionization) of $\gamma'$ with heavy mirror
metal components, which is possible because these components 
retain their K-shell mirror electrons.
As far as the heating of the halo is concerned,
it might be sufficient (at least as a rough approximation) to include just the
mirror iron ($Fe'$) component 
provided that the proportion of the supernova $\gamma'$ energy 
contributed by $\gamma'$ with $E_{\gamma'}$
less than the $Fe'$ K-shell binding energy, $I \approx$ 9 keV,
is small. 
[Although we consider a mirror metal component consisting of just $Fe'$,
reasonable alternatives e.g. replacing $Fe'$ with mirror oxygen ($O'$)
do not significantly affect any of our results.]

The total photoelectric cross-section\footnote{Unless otherwise indicated, 
we use natural units with $\hbar = c = 1$.}
of $Fe'$ ($Z = 26$) is given by (see e.g. \cite{book5}):
\begin{eqnarray}
\sigma_{PE} (E_{\gamma'}) = 
{g 16\sqrt{2} \pi \over 3m_e^2 } \alpha^6  Z^5 \left[ {m_e \over E_{\gamma'}
}\right]^{7/2} 
\ \ {\rm for} \ E_{\gamma'} \gg I
\ .
\label{pe4}
\end{eqnarray}
Here $g = 1$ or 2 counts the number 
of K-shell mirror electrons present.
The $Fe'$ number density can be parameterized in terms of the halo $Fe'$ mass fraction, $\xi_{Fe'}$:  
\begin{eqnarray}
n_{Fe'} = n_{He'} \left(1 + {f \over 4}\right) \left({m_{He} \over m_{Fe}}\right) \left({\xi_{Fe'} \over 1 - \xi_{Fe'}}\right)
\ 
\end{eqnarray}
with $ 
f\equiv n_{H'}/n_{He'} = 0.4$ suggested by early Universe cosmology \cite{paolo1}.

The flux, $F(r)$, of supernova $\gamma'$ at a particular point, $P$, will deposit an energy per unit volume per unit time
of: 
\begin{eqnarray}
{d^2 E_{in} \over dt dV}=  \int {dF (r) \over dE_{\gamma'}} \ n_{Fe'}  (r) \ \sigma_{PE} 
\ dE_{\gamma'}
\ .
\label{meal}
\end{eqnarray} 
If the (average) supernova flux traces the stellar mass density, $\rho_D (r)$,
assumed spherically symmetric for simplicity then:
\begin{eqnarray}
{dF(r) \over dE_{\gamma'}} = 
R_{SN} \ E_{\gamma'} {dN_{\gamma'} \over dE_{\gamma'}} 
\ 
\int^{\infty}_0 \int^1_{-1} \ {\rho_D \over m_D} \ {e^{-\tau} \ r'^2 \over 2{\rm d}^2}\ \ d\cos\theta dr'
\ .
\label{33a}
\end{eqnarray}
Here,
$R_{SN}$ is
the frequency of type II supernova in the galaxy under consideration,
${\rm d} = \sqrt{ r^2 + r'^2 - 2rr' \cos\theta}$ is the distance from the supernova to the point $P$ and 
$\tau$ is the  optical depth along this path: $\tau = \int_0^{\rm d} n_{Fe'} \sigma_{PE} \ dy $.
We define the
spherically symmetric stellar mass density by
requiring that the mass within a radius $r$ is the same as that of the 
Freeman disk, with surface density
$\Sigma_* = {m_D \over 2\pi r_D^2} \ e^{-r/r_D}$. This implies that 
\begin{eqnarray}
\rho_D (r) = {m_D \over 4\pi r_D^2 r} \ e^{-r/r_D} \ .
\end{eqnarray}
Here $m_D$ is the total stellar mass of the disk and $r_D$ is the disk scale length.

An important ingredient is the frequency of supernova in a given galaxy, $R_{SN}$.
In our previous study \cite{foot5} we used the rough
baryonic scaling relation \cite{sal3} $m_D \propto (L_B)^{1.3}$ and $R_{SN}\propto (L_B)^{0.73}$
from the supernova study \cite{sn}.
These relations together imply $R_{SN} \propto (m_D)^{0.56}$ for spiral galaxies.
The supernova study \cite{sn} also provides a measurement of $R_{SN}$ versus $\stackrel{\sim}{m}_D$,
where $\stackrel{\sim}{m}_D$ is the stellar mass 
derived from photometry and spectral fitting,
finding $R_{SN} \propto (\stackrel{\sim}{m}_D)^{0.45}$.
In this work we take $R_{SN} \propto (m_D)^{0.5}$. Of course, the systematic uncertainty is significant, later
we examine the effect of a variation of $\pm 0.15$ in the exponent of this relation.

To proceed, we can parameterize the  
mirror photon energy spectrum from an (average) single supernova
via a power law, with a cut-off at $E_{\gamma'} = E_c$:
\footnote{
Of course, the $\gamma'$ spectrum will not be a simple power law over all energies, however
only the part of the spectrum with $E_{\gamma'} \stackrel{<}{\sim} 30$ keV   is important since
galaxy halos are optically thin for $\gamma'$ with energies greater than around 30 keV \cite{foot5}.}
\begin{eqnarray}
E_{\gamma'} {dN_{\gamma'} \over dE_{\gamma'}}  
\equiv  \kappa \left(E_{\gamma'}\right)^{c_1}
\ 
.
\label{m3}
\end{eqnarray}
With this parameterization, $R_{SN} \int_0^{E_c} \kappa \left(E_{\gamma'}\right)^{c_1} \ dE_{\gamma'} \equiv  
\ {L'}_{SN}$
is the total $\gamma'$ energy produced (on average) by ordinary supernova per unit time in the galaxy
under consideration.
With the above definitions, and the assumed scaling, $R_{SN} \propto (m_D)^{0.5}$, we have:
\begin{eqnarray}
\kappa\ R_{SN} = {1 + c_1 \over (E_c)^{1+c_1}}
\left( {m_D \over m_D^{MW}}\right)^{0.5}   
\ {L'}_{SN}^{MW}
\label{kap2}
\end{eqnarray}
where ${L'}_{SN}^{MW}$ is the total $\gamma'$ energy produced (on average) by ordinary supernovae for a
reference $\sim $ Milky Way sized spiral galaxy of stellar mass $m_D^{MW} = 5\times 10^{10} \ m_\odot$. 

\vskip 0.7cm
\noindent
{\it (c) hydrostatic equilibrium}
\vskip 0.3cm
\noindent

Both the heating and cooling depend on the halo temperature, $T(r)$. The halo temperature can be
evaluated assuming hydrostatic equilibrium,
where the force of gravity is balanced by
the pressure gradient. That is,
\begin{eqnarray}
{dP \over dr} &=& -\rho (r) g(r) 
\ .
\label{p9}
\end{eqnarray}
Here, $\rho (r) = \bar m n_T (r)$ where
$n_T (r)$ is the number density of the plasma mirror particle component
\footnote{As in our previous study \cite{foot5},  we neglect a possible small
dark disk/compact object component made of old mirror stars, mirror white dwarfs etc,
so that dark matter consists only of the plasma component.}
with mean mass, $\bar m$.
For a fully ionized plasma, arguments from early Universe cosmology suggest that $\bar m \simeq 1.1$ GeV \cite{paolo1}.
The local acceleration due to gravity, $g(r)$, is given in terms of Newton's constant, $G_N$:  
\begin{eqnarray}
g(r) = {v_{rot}^2 \over r} \equiv {G_N \over r^2} \int_0^r \left[ \rho (r)  + \rho_{baryon}(r)\right] \ dV
\ .
\end{eqnarray}
Although in principle, the baryonic density contains both a stellar and gas component,  
in this work, we make the simplifying approximation of considering only the stellar mass contribution to
$\rho_{baryon}$. 
This can be justified as follows: The gas component of spirals is typically smaller than the stellar
component and importantly its distribution is more (radially) extended.
Thus at any given radius, the gas contribution to $g(r)$ is generally much
smaller than either the stellar contribution or the dark matter contribution.

To solve the hydrostatic equilibrium equation, we need to assume a boundary condition.
We assume the boundary condition $dT/dr \to 0$ at large galactic radius, $R_{gal}$ (taken to be
50$r_D$). Our numerical results are
independent, to a very good approximation, of the particular value of $R_{gal}$ chosen 
so long as $R_{gal} \gg r_D$.

The conditions, Eq.(\ref{meal8},\ref{p9}), can be solved numerically. The strategy employed 
in our previous work \cite{foot5}
was to assume that the dark matter profile had the Burkert form: \cite{bp}
\begin{eqnarray}
\rho (r) =  {\rho_0 r_0^3  \over (r^2 + r_0^2) (r + r_0)
} 
\ .
\label{1}
\end{eqnarray}
We then obtained $r_0, \ \rho_0$ by numerically minimizing the function:
\begin{eqnarray}
\Delta (r_0, \rho_0) \equiv {1 \over 10 r_D} \int^{11r_D}_{r_D}
\left| 1 - {{d^2 E_{in} \over dt dV} \over {d^2 E_{out} \over dt dV}}\right| \ dr
\ . 
\label{delta}
\end{eqnarray}
It was found \cite{foot5}
that the Burkert profile provided a rough solution to the dynamical
condition, Eq.(\ref{meal8}), with minimum value of $\Delta$ around 0.1 (roughly independently of $m_D$ 
in the range of interest: $10^{9} \ m_\odot \stackrel{<}{\sim} m_D \stackrel{<}{\sim} 10^{12} \ m_\odot$).
That is, the left and right-hand sides of Eq.(\ref{meal8}) agree to within
about 10 \%.

The main reason the Burkert profile was adopted in \cite{foot5} was to compare
the derived values of $r_0, \rho_0$ with the corresponding values obtained by Salucci and others \cite{core}
which assumed that profile.
In this work, we adopt a more general approach and assume that the dark matter
distribution takes the form:
\begin{eqnarray}
\rho (r) =  \rho_0 \left[ {r_0^2  \over r^2 + r_0^2
}\right]^\beta
\ .
\label{beta}
\end{eqnarray} 
It turns out that this more general profile allows a near exact solution to the  
condition, Eq.(\ref{meal8}). The resulting $\Delta$ minimum obtained
by varying $r_0, \ \rho_0$ and now $\beta$ is less than 0.01.
That is, the left and right-hand sides of Eq.(\ref{meal8}) agree to better than
1\%
\footnote{
We have also checked that $\Delta_{min} \stackrel{<}{\sim} 0.01$ holds true even when the
range of integration of the integral in Eq.(\ref{delta}) is extended from $\{r_D \le r \le 11 r_D\}$ to $\{0.2r_D \le r \le  15 r_D\}$.
Extending this integration range also has negligible effect on the derived values for $\beta, r_0$ and $\rho_0$.
}.
We expect therefore that this profile, and the values of
$r_0, \ \rho_0$ and $\beta$
derived by minimizing $\Delta$, should provide an accurate representation
of the dark matter properties expected from this dynamics.  

An empirical baryonic scaling relation
is known which
relates $m_D$ to $r_D$ for spiral galaxies \cite{lapi,PSS}:
\begin{eqnarray}
\log \left( {r_D \over {\rm kpc}}\right) \approx 0.633 + 0.379
\log \left( {m_D \over 10^{11} \ m_{\odot}}\right) 
+ 0.069 \left[\log \left( {m_D \over 10^{11} \ m_{\odot}}\right)\right]^2
\ .
\label{rd}
\end{eqnarray} 
With this relation, the baryonic parameters of spirals are (roughly) specified by a single
parameter which can be taken as either  $m_D$ or  $r_D$.

To continue, we must choose values of ${L'}_{SN}^{MW}, \ c_1, \ \xi_{Fe'}$ and then we can
numerically solve the equations, minimizing $\Delta (\rho_0, r_0, \beta)$ to give
values of $\rho_0, r_0$ and $\beta$ for a given spiral galaxy parameterized by $m_D, \ r_D$.
We consider stellar disk masses in the range,
$10^{9} \ m_\odot \stackrel{<}{\sim} m_D \stackrel{<}{\sim} 10^{12} \ m_\odot$, and parameterize $r_D$ via
Eq.(\ref{rd}). 
This mass range covers the typical $m_D$ values for spiral galaxies. 
The result of performing this numerical task is the following. We find that $\Delta_{min} \stackrel{<}{\sim} 0.01$ (independently
of $m_D$) and 
\begin{eqnarray}
\beta & \simeq & 1.0 \nonumber \\
r_0 & \simeq & 1.4 \left( {r_D \over {\rm kpc}}\right)   \ {\rm kpc} \nonumber \\
\rho_0 r_0 &\simeq & \left[ {\xi_{Fe'} \over 0.02}\right]^{0.8} \ 
\left[ { {L'}_{SN}^{MW} \over 10^{45} \ {\rm erg/s}} \right]^{0.8}
\left[ {2 \over c_1}\right] 
\ 50 \ m_\odot/{\rm pc}^2
\ .
\label{40}
\end{eqnarray}
Evidently, this dissipative halo dynamics yields a quasi-isothermal dark matter
density profile,
a result that is not unexpected given earlier analytic work of ref. \cite{sph,foot13}.
In particular, it was shown in those references that the behaviour: $\rho (r) \propto 1/r^2$ leads to 
an approximate solution to the hydrostatic and energy balance equations for $r \gg r_D$
(where the supernova energy source can be modelled as a point source and the matter density is dominated by mirror
dark matter). For $r \stackrel{<}{\sim} few \ r_D$, the distribution of the supernova energy source over a
finite volume weakens the heating rate, and consequently, energy balance requires a cored halo \cite{foot13}.

We emphasize that the first two results in Eq.(\ref{40}) above, hold even when the parameters (${L'}_{SN}^{MW}, \
 \xi_{Fe'}, \ c_1$) are varied.
This is illustrated in figure 1a,b,c, where we give the results for $\beta, \ r_0, \ \rho_0 r_0$ obtained
from minimizing $\Delta$ for fixed    
$c_1 = 2$, $\xi_{Fe'} = 0.02$, $E_c = 50$ keV but 
with an order-of-magnitude variation in ${L'}_{SN}^{MW}$. [Alternatively
fixing ${L'}_{SN}^{MW}$ and varying $E_c$ is
equivalent as both operations simply scale the supernova energy $\gamma'$ flux.]  
Similar
results hold when $\xi_{Fe'}$ is varied and are therefore not shown.
Variation of $c_1$ also has little effect for the first two relations in Eq.(\ref{40}).
Note that the derived
scaling of the core radius with disk scale length is consistent
with observations \cite{dsds}.
The third relation,
the one for $\rho_0 r_0$, does depend on the parameters.  
However there is an approximate parameter degeneracy, since 
this quantity depends on only one combination of these parameters. 
Also, note that these parameters are likely to be approximately independent
of galaxy size, so that a $\rho_0 r_0 \propto constant$ scaling 
is expected. Again such a scaling relation is consistent with observations \cite{core}. 

\vskip 0.2cm
\centerline{\epsfig{file=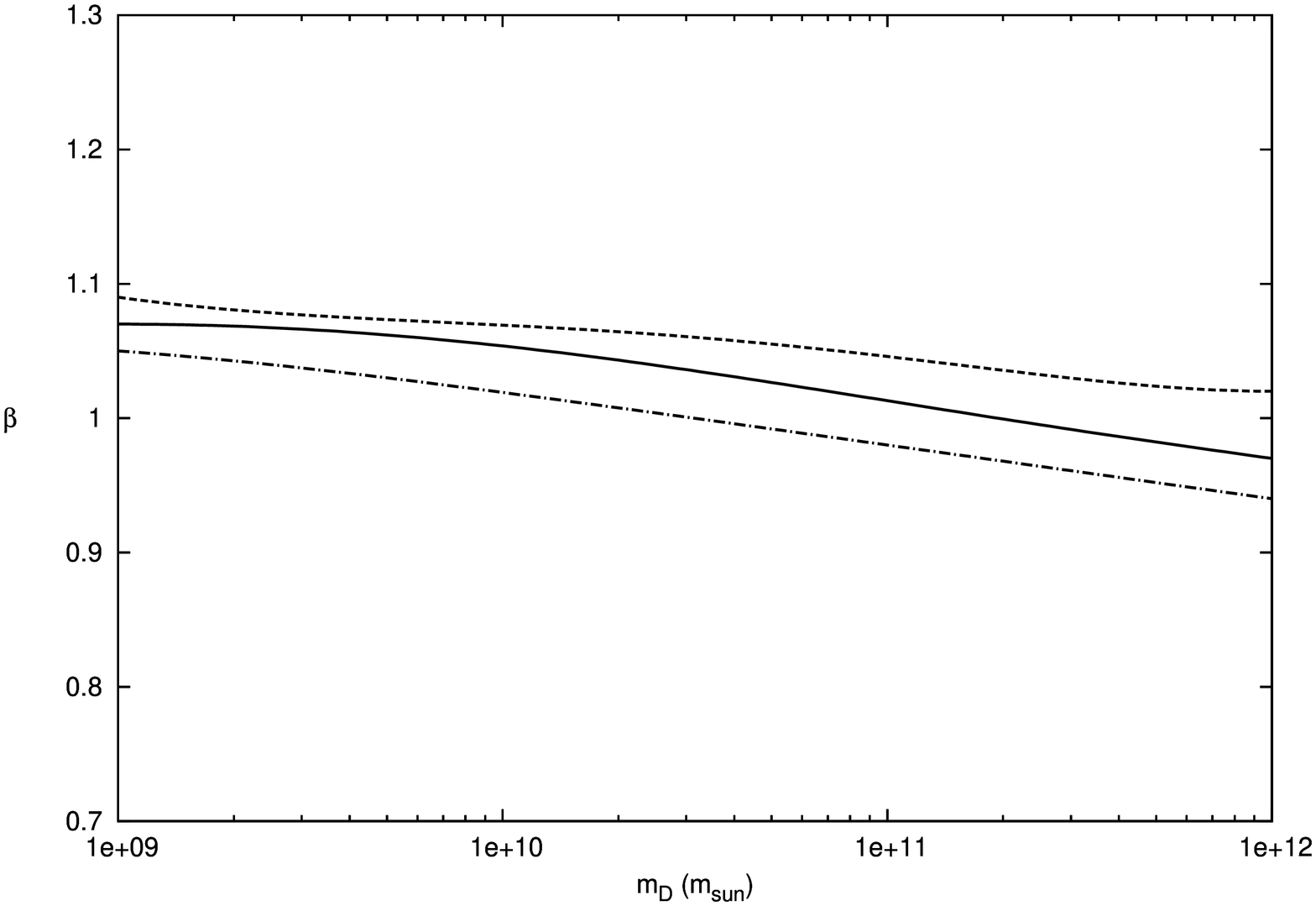, angle=270, width=13.2cm}}
\vskip 0.2cm
\noindent
{\small Figure 1a: 
The dark matter density, slope, $\beta$ [Eq.(\ref{beta})], versus $m_D \ [m_\odot]$.
The dashed-dotted, solid and dotted lines correspond to 
${L'}_{SN}^{MW} = 0.3\times 10^{45}$ erg/s, 
${L'}_{SN}^{MW} = 1.0\times 10^{45}$ erg/s and 
${L'}_{SN}^{MW} = 3.0\times 10^{45}$ erg/s respectively.  
}

\vskip 0.8cm
\centerline{\epsfig{file=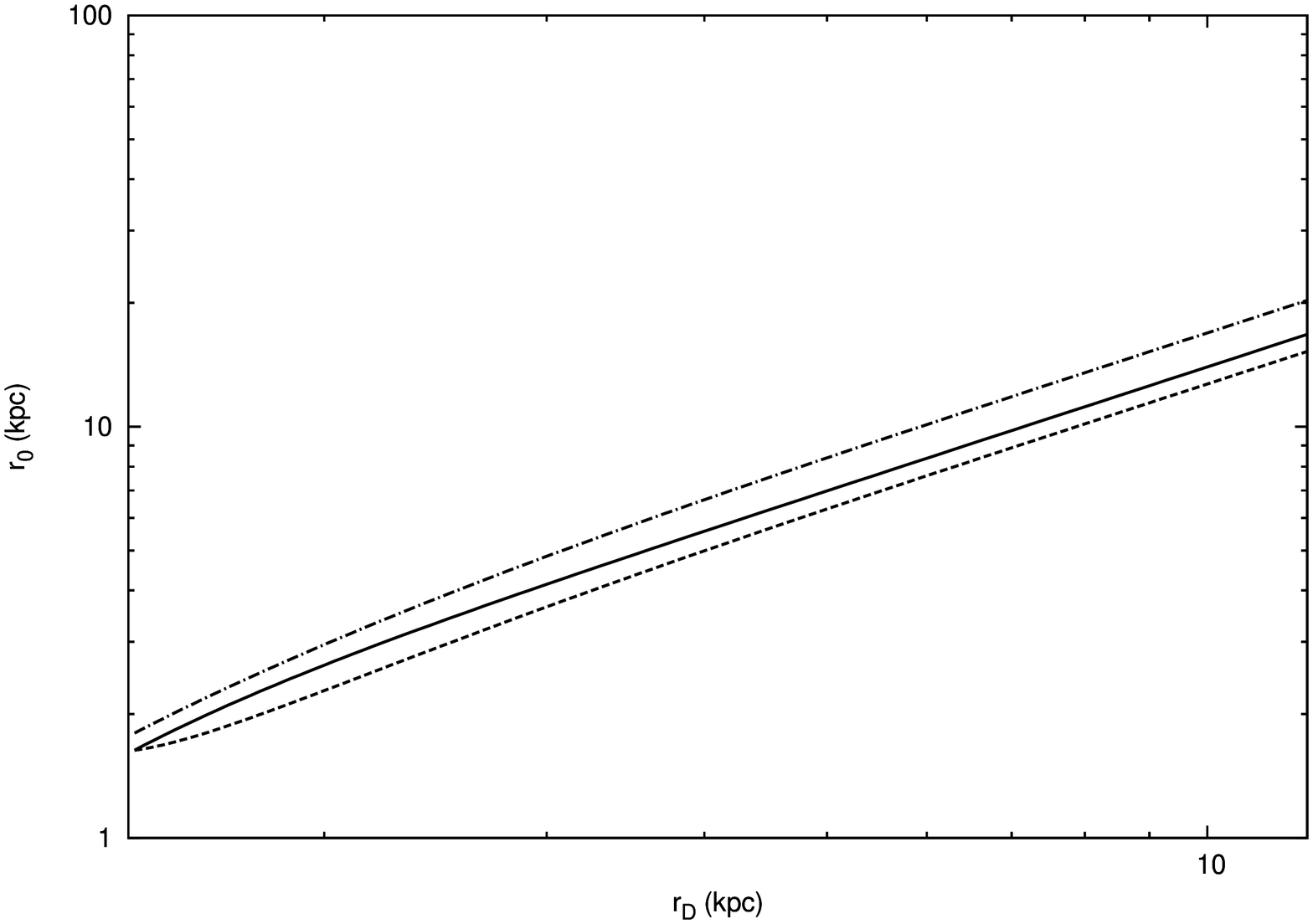, angle=270, width=13.2cm}}
\vskip 0.2cm
\noindent
{\small Figure 1b: 
Dark matter core radius, $r_0$, versus
disk scale length, $r_D$. 
Parameters as per figure 1a.
}
\vskip 0.4cm

\centerline{\epsfig{file=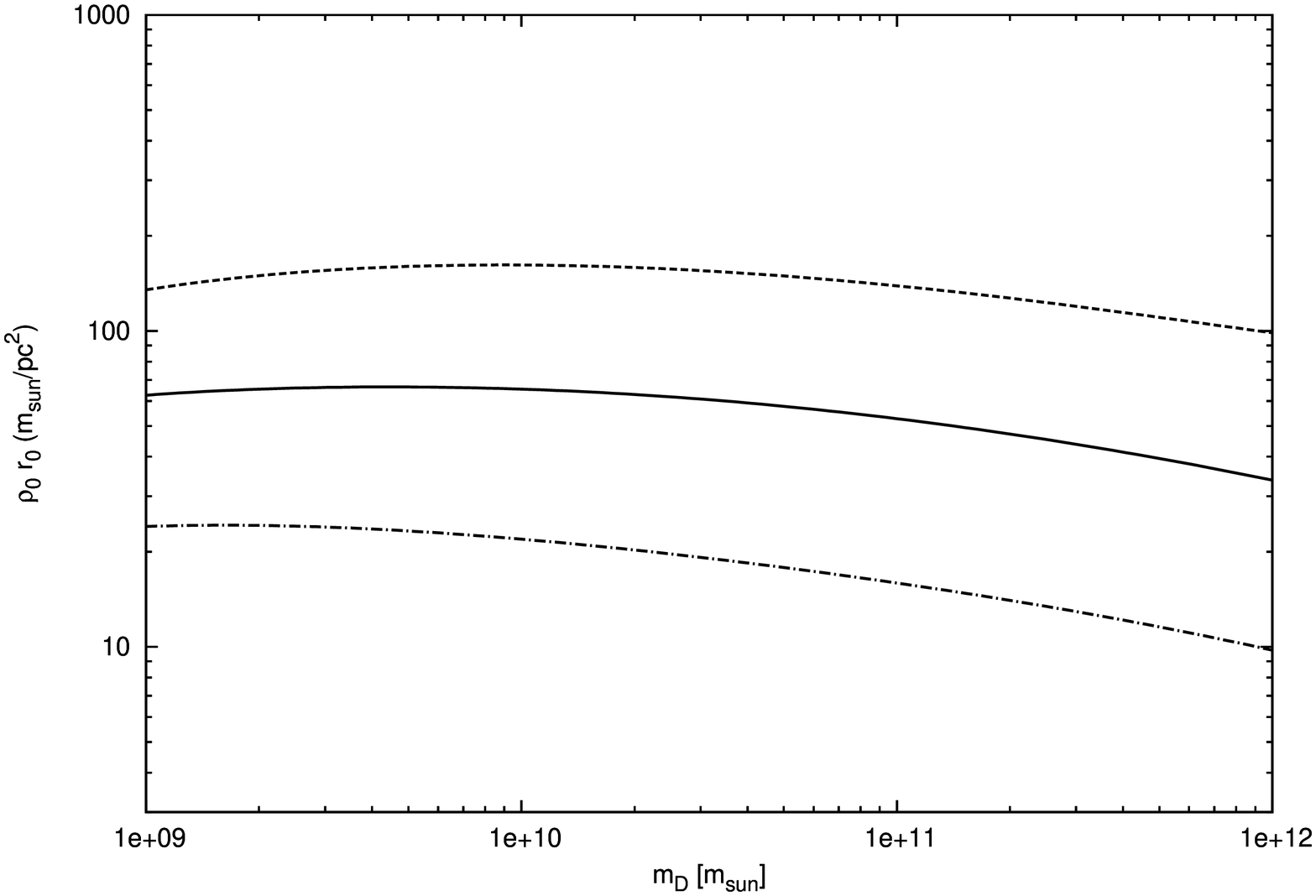, angle=270, width=13.2cm}}
\vskip 0.2cm
\noindent
{\small Figure 1c: 
$\rho_0 r_0$ versus $m_D$.
Parameters as per figure 1a.
}
\vskip 1cm

The equations governing dissipation and heating depend on the ionization state
of the halo. In particular the number of $Fe'$ K shell states occupied [the $g$ factor in Eq.(\ref{pe4})] depends on the ionization
state of $Fe'$ while the ion density in
Eq.(\ref{john}) depends on the ionization state of $H',\ He'$.
The equations governing the ionization state have been given in \cite{foot5}
and depend only on the temperature of the mirror particle plasma.
[These equations are, of course, included in our numerical work here.]
Interestingly we find that the  mean plasma temperature 
corresponding to the $m_D$ range  
$10^{9} \ m_\odot \stackrel{<}{\sim} m_D \stackrel{<}{\sim} 10^{12} \ m_\odot$ is
approximately
$20\ {\rm eV} \stackrel{<}{\sim} \ T \ \stackrel{<}{\sim} {\rm keV} $. 
This is roughly the range for which (a) $H'$, $He'$ are typically fully ionized and
(b) $Fe'$ has both K-shell states filled.  
This consistency of the ionization state of the halo, over the entire mass range of
interest,
$10^{9} \ m_\odot \stackrel{<}{\sim} m_D \stackrel{<}{\sim} 10^{12} \ m_\odot$ 
is necessary to obtain the smooth
behaviour of the derived relations, Eq.(\ref{40}).
This would appear to be an important constraint if one were to think about replacing
mirror dark matter with a more generic dissipative hidden sector model.

In figure 2 we plot the evaluated temperature versus radial distance for some examples. 
For each of these examples the equation governing hydrostatic 
equilibrium, Eq.(\ref{p9}), is solved with
$\beta, \ r_0$ and $\rho_0$ obtained
by minimizing $\Delta$, Eq.(\ref{delta}) [with reference parameters: 
$c_1 = 2, \ \xi_{Fe'} = 0.02$, ${L'}_{SN}^{MW} = 10^{45}$ erg/s, 
$E_c = 50$ keV are assumed].  

\vskip 0.8cm
\centerline{\epsfig{file=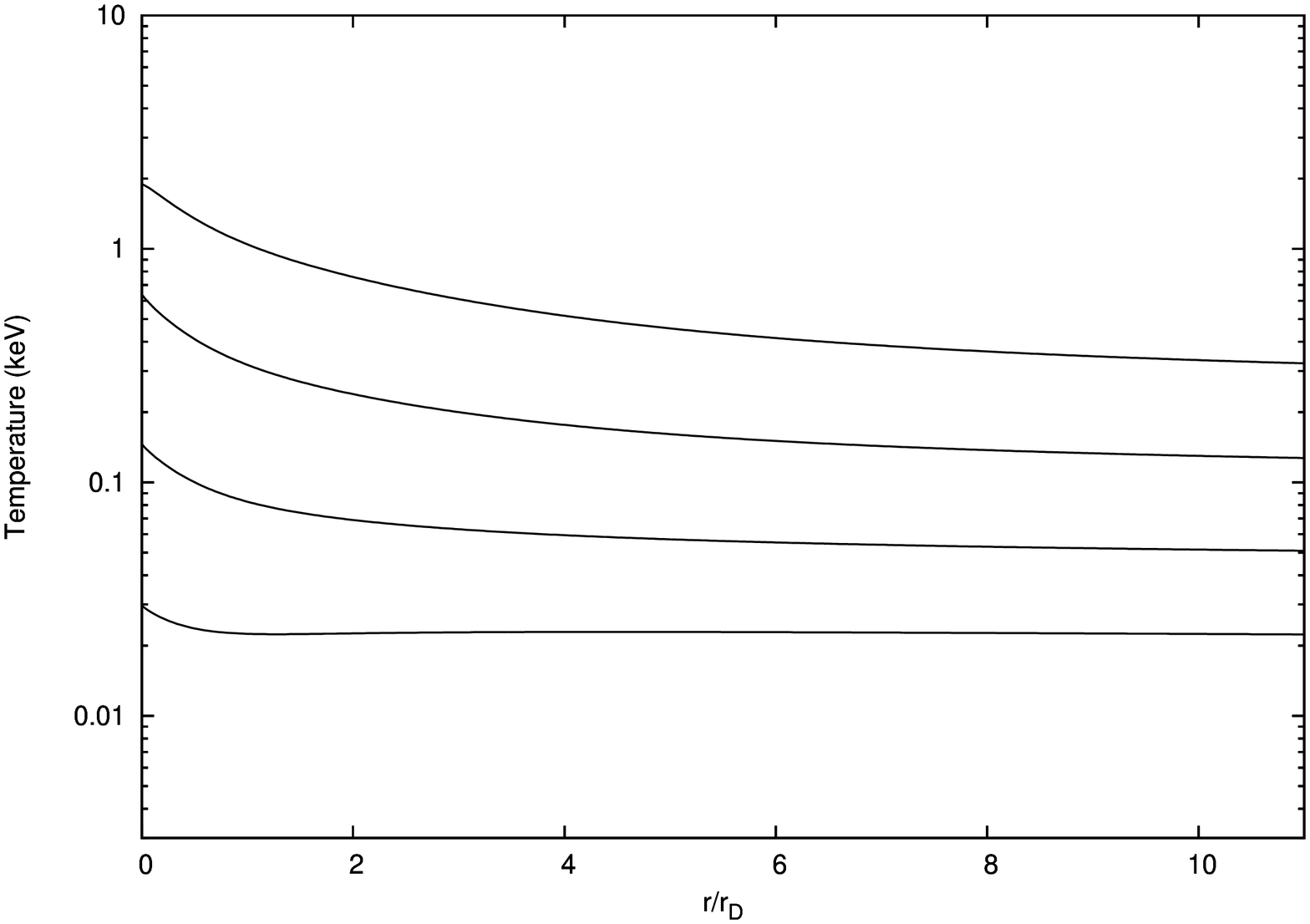, angle=270, width=13.2cm}}
\vskip 0.3cm
\noindent
{\small Figure 2: Halo mirror plasma temperature versus $r/r_D$ for
the examples with 
(from bottom to top curves): $m_D = 10^9 \ m_\odot, \ m_D = 10^{10} \ m_\odot, \ m_D = 10^{11} \ m_\odot, \
m_D = 10^{12} \ m_\odot$.
}

\vskip 1cm

Examples of the rotation curves predicted by this dynamics are given in figures 3,4.
Consider first a specific example, for which we choose the galaxy NGC3198.
This galaxy has stellar mass around
$m_D = 3.0 \times 10^{10}\ m_\odot$ \cite{blok} and from Eq.(\ref{rd}) we find $r_D = 2.8$ kpc.
We use the measurement of the rotation curve from \cite{ngc3198} which is consistent
with other measurements such as the one in \cite{blok}.
In figure 3 we give our result for the rotation curve, determining the dark matter parameters, $\beta,\  r_0, \ \rho_0$
by minimizing $\Delta$ inputting the above baryonic parameters for NGC3198.
The we that that the data could be roughly fit with the parameters 
taken to be: 
$c_1 = 2, \ \xi_{Fe'} = 0.02$, ${L'}_{SN}^{MW} = 2.2\times 10^{45}$ erg/s and 
$E_c = 50$ keV. 
In figure 4 we show derived rotation curves obtained for representative examples with 
(from bottom to top curves)
$m_D = 10^9 \ m_\odot, \ 
m_D = 10^{10}  \ m_\odot, \ \ m_D = 3\times 10^{10} \ m_\odot,\
m_D = 10^{11}  \ m_\odot, \ \ m_D = 3\times 10^{11} \ m_\odot$.
The reference  parameters taken are the same as per figure 2.

\vskip 0.6cm 

\centerline{\epsfig{file=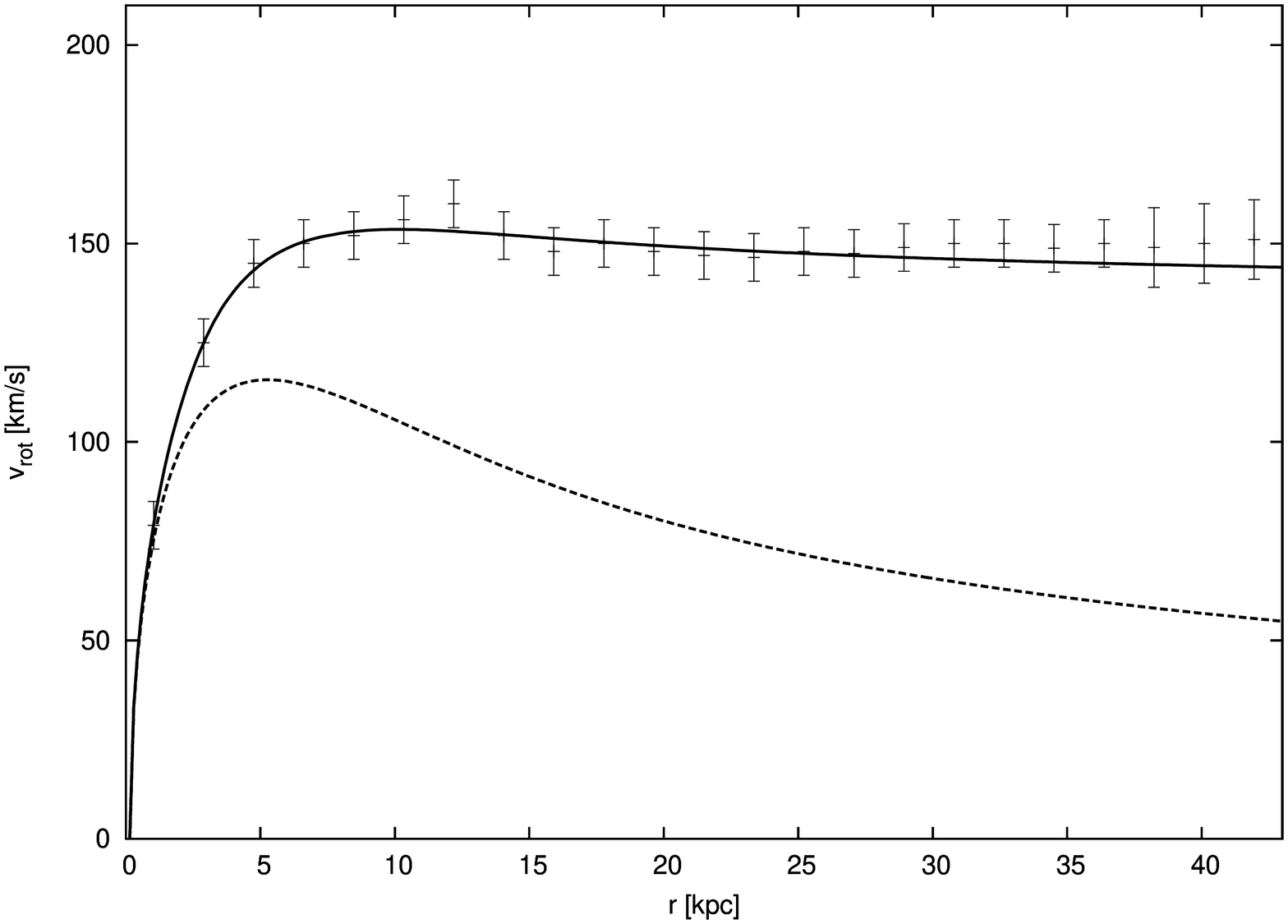, angle=270, width=13.2cm}}
\vskip 0.3cm
\noindent
{\small Figure 3: Rotation curve for NGC3198. The solid line is the `theoretical' rotation curve derived
from the assumed halo dynamics.  Also shown (dashed curve) is the baryonic contribution.
The data is obtained from \cite{ngc3198}.
}
\vskip 0.8cm

\centerline{\epsfig{file=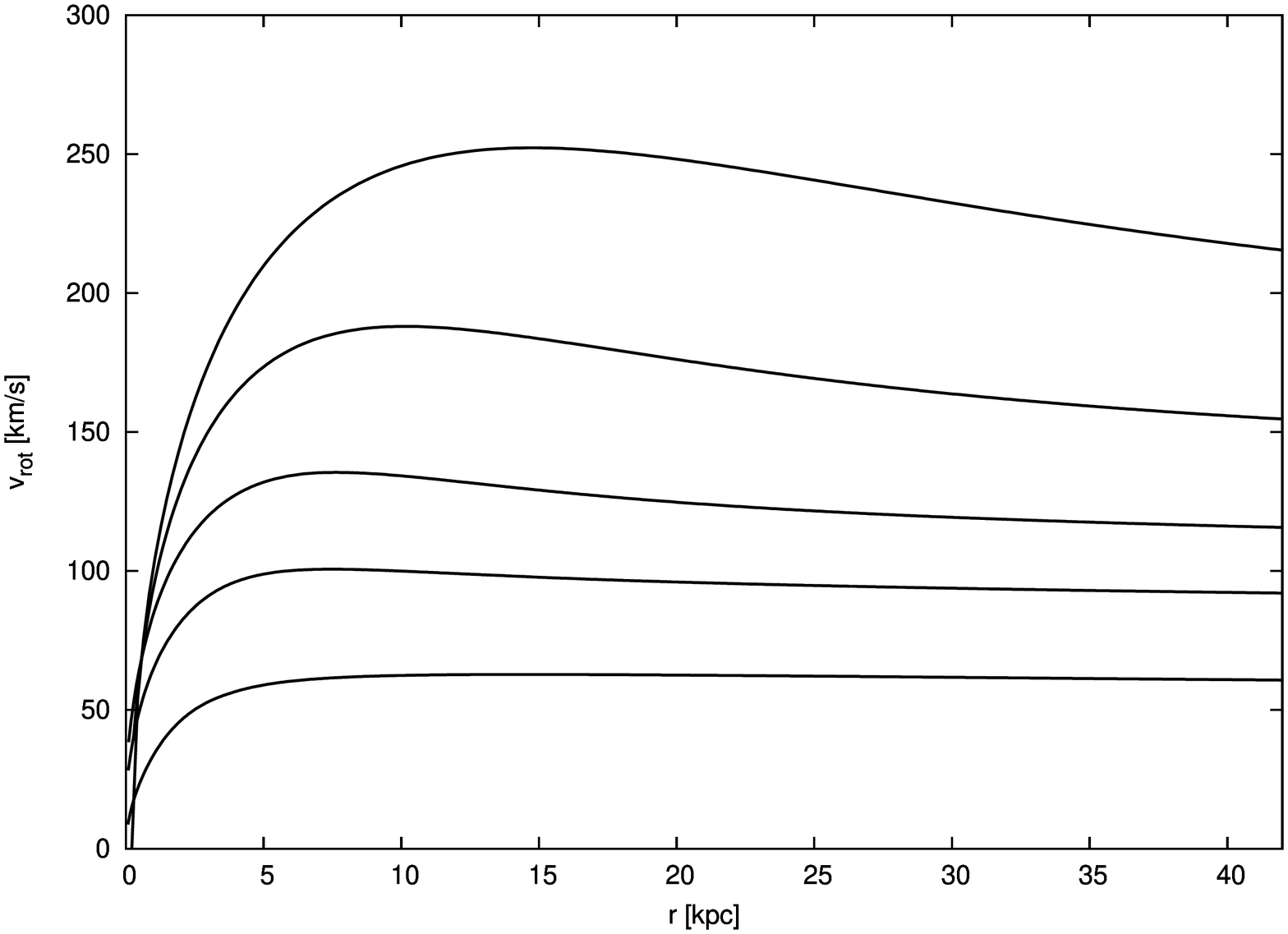, angle=270, width=13.2cm}}
\vskip 0.3cm
\noindent
{\small Figure 4: Rotation curves for examples with
$m_D = 10^9 \ m_\odot, \ 
m_D = 10^{10}  \ m_\odot, \ \ m_D = 3\times 10^{10} \ m_\odot, \
m_D = 10^{11}  \ m_\odot, \ \ m_D = 3\times 10^{11} \ m_\odot$.
}

\vskip 1cm

Tully and Fisher discovered some time ago that
the luminosity  of a spiral galaxy has a tight relation
with the maximum value
of its rotational velocity \cite{tf}.
Current estimates, e.g. \cite{bell,t1,mnrs}, indicate that $L_B \propto v_{max}^{\alpha_1}$, with
$\alpha_1 \approx 3.0-3.5$ ($L_B$ is the B-band luminosity). A baryonic Tully-Fisher 
relation is also known to approximately relate $m_D \propto v_{max}^{\alpha_2}$ 
with $\alpha_2 \approx 4.0-4.5$.
Given that we can derive the dark matter profile via the assumed halo dynamics
we can also work out the $L_B$ versus $v_{max}$ dependence.  
To do this, we  
obtain $L_B$ from the measured scaling, $R_{SN} \propto (L_B)^{0.73}$ \cite{sn}. 
As before we minimize the function $\Delta$ considering first     
the usual reference parameters:
${L'}_{SN}^{MW} = 10^{45}$ ergs/s,
$c_1 = 2$, $\xi_{Fe'} = 0.02$, $E_c = 50$ keV. 
The result of this numerical work is shown in figure 5 for $L_B$ versus $v_{max}$ and
figure 6 for $m_{D}$ versus $v_{max}$ .

Our numerical results can be excellently approximated by a power law:
$L_B \propto v_{max}^{\alpha_1}$ and $m_{D} \propto v_{max}^{\alpha_2}$,
with $\alpha_1 \simeq 2.9$ ($\alpha_2 \simeq 4.1$).
We have also investigated the dependence on the slope parameter on variation of
our parameters: ${L'}_{SN}^{MW}, \ \xi_{Fe'}, \ c_1$. Considering $\alpha_1$ (similar results
hold for $\alpha_2$) we find that an order-of-magnitude variation in 
${L'}_{SN}^{MW}$ or $\xi_{Fe'}$ 
around the
reference value changes the slope $\alpha_1$ by $\pm 0.4$ (with larger values of 
${L'}_{SN}^{MW}$ or $\xi_{Fe'}$ steepening the slope). 
Variation of $c_1$ over the range: $1 \le c_1 \le 3$ modifies $\alpha_1$ by $\pm 0.2$. 
We also considered the effect of a possible variation of the exponent, 
$\delta$, in the relation $R_{SN} \propto (m_D)^{\delta}$,
over the range $\delta = 0.50 \pm 0.15$. Variation of $\delta$ over this range affects the slope parameter $\alpha_1$ by 
$\pm 0.4$.  
Of course, the consistency of the derived slope parameter with the value $\alpha_1 \approx 3$ 
inferred from observations is 
not so surprising given 
that the galactic scaling relations,
derived in part with the same types of data, were found to be satisfied with
this halo dynamics in \cite{foot5}. Nevertheless, working directly with 
relations such as those of Tully and Fisher  might be useful to further scrutinize
this dynamical halo model. 

\vskip 0.3cm

\centerline{\epsfig{file=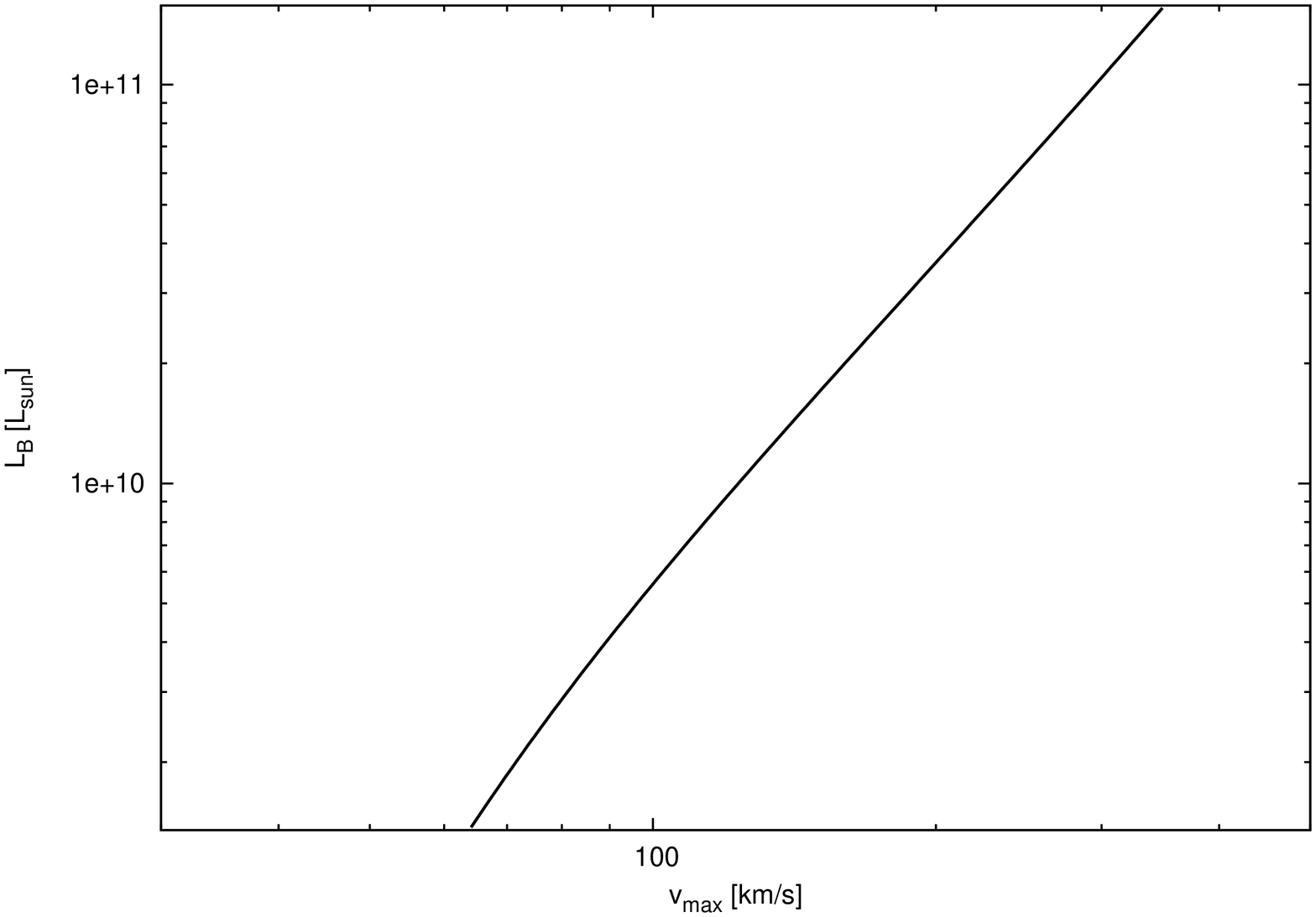, angle=270, width=13.2cm}}
\vskip 0.3cm
\noindent
{\small Figure 5: $L_B$ versus $v_{max}$, with halo profile derived 
from dissipative mirror dark matter.
}
\vskip 1.0cm
\centerline{\epsfig{file=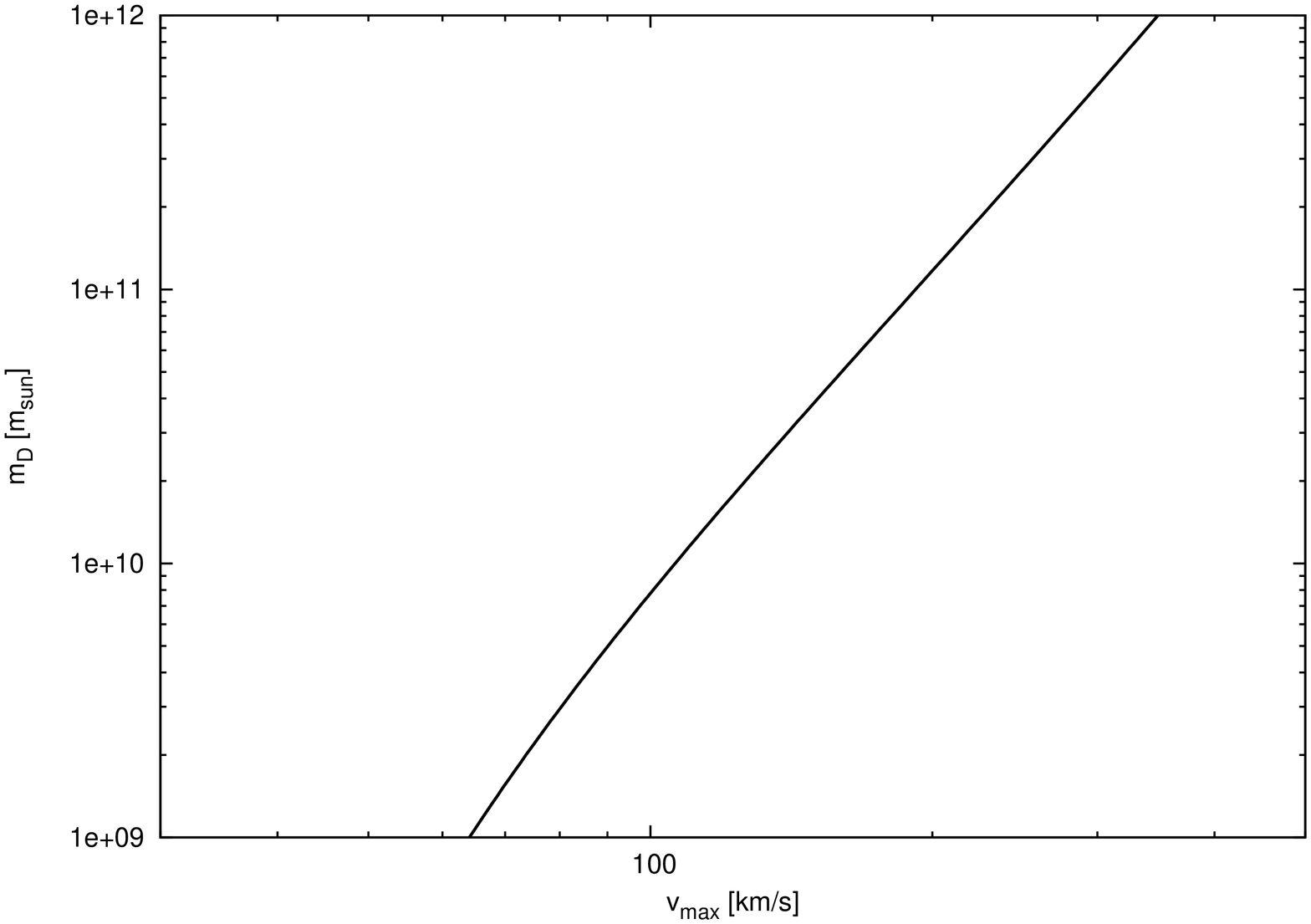, angle=270, width=13.2cm}}
\vskip 0.3cm
\noindent
{\small Figure 6: Stellar disk mass $m_{D}$ versus $v_{max}$, with halo profile derived 
from dissipative mirror dark matter.
}
\vskip 1.0cm

To summarize,
galaxy structure has presented a fascinating puzzle for many years. 
Dissipative dark matter candidates such as mirror dark matter offer an
interesting approach to a possible solution.
In this picture the halo has nontrivial dynamics: The energy lost due
to dissipation is replaced ultimately by heating from ordinary supernovae. 
Such a mechanism requires small kinetic mixing interaction, $\epsilon \sim 10^{-9}$,
for which there is independent evidence from direct detection experiments \cite{footrec}.
The heating from ordinary supernovae provides a direct coupling between the
dark matter and ordinary matter in galaxies. 
In the framework of mirror dark matter such a dark matter picture is quite predictive:
the dark matter distribution, $\rho (r)$,  can be completely determined from the baryonic properties
of galaxies. We found that 
$\rho (r) \simeq \rho_0 r_0^2/(r^2 + r_0^2)$, where
the
core radius, $r_0$, scales with disk scale length, $r_D$, via
$r_0/{\rm kpc} \approx 1.4\left(r_D/{\rm kpc}\right)$.
Additionally,  the product $\rho_0 r_0$ is roughly
$constant$, i.e.
independent of galaxy size
(the $constant$ is set by the parameters of the model).
We further show that such a dark matter distribution leads to rotation curves which
satisfy the Tully-Fisher relation.

\vskip 1.0cm
\noindent
{\large \bf Acknowledgements}

\vskip 0.2cm
\noindent
The author would like to thank Filippo Mannucci for very helpful
correspondence.
This work was supported by the Australian Research Council.


\begin{thebibliography}{999}

\bibitem{sph}
R. ~Foot and R. ~R. ~Volkas,
Phys. Rev. D{\bf 70}, 123508 (2004) [astro-ph/0407522].

\bibitem{review}
Z.~K.~Silagadze,
Acta Phys.\ Polon.\ B {\bf 32}, 99 (2001) [hep-ph/0002255];
A.~Y.~Ignatiev and R.~R.~Volkas,
hep-ph/0306120;
R.~Foot,
Int.\ J.\ Mod.\ Phys.\ D {\bf 13}, 2161 (2004) [astro-ph/0407623];
Int.\ J.\ Mod.\ Phys.\ A {\bf 19}, 3807 (2004)
[astro-ph/0309330];
Z.~Berezhiani,
Int.\ J.\ Mod.\ Phys.\ A {\bf 19}, 3775 (2004) [hep-ph/0312335];
P.~Ciarcelluti,
Int.\ J.\ Mod.\ Phys.\ D {\bf 19}, 2151 (2010)
[arXiv:1102.5530].

\bibitem{review2}
R.~Foot,
Int.\ J.\ Mod.\ Phys.\ A {\bf 29}, 1430013 (2014)
[arXiv:1401.3965].





\bibitem{foot5}
R.~Foot,
arXiv:1304.4717. 

\bibitem{flv}
R. ~Foot, H. ~Lew and R. ~R. ~Volkas, Phys. Lett. B{\bf 272}, 67 (1991);
Mod. Phys. Lett. A{\bf 7}, 2567 (1992);
R.~Foot and R.~R.~Volkas,
Phys.\ Rev.\ D {\bf 52}, 6595 (1995) [hep-ph/9505359].

\bibitem{he}
R. ~Foot and X-G. ~He, Phys. Lett. B{\bf 267}, 509 (1991). 

\bibitem{holdom} 
B.~Holdom,
Phys.\ Lett.\ B {\bf 166}, 196 (1986).


\bibitem{bookastro}
G. ~B. ~Rybicki and A.~P. ~Lightman,
``Radiative processes in astrophysics'',
{\it Wiley, (2008)}.


\bibitem{raffelt}
G.~G.~Raffelt,
``Stars As Laboratories For Fundamental Physics: The Astrophysics Of
Neutrinos, Axions, And Other Weakly Interacting Particles,''
{\it  Chicago, USA: Univ. Pr. (1996) 664 p};
S.~Davidson, S.~Hannestad and G.~Raffelt,
JHEP {\bf 0005}, 003 (2000) [hep-ph/0001179];
R.~N.~Mohapatra and I.~Z.~Rothstein,
Phys.\ Lett.\ B {\bf 247}, 593 (1990).

\bibitem{sil}
R.~Foot and Z.~K.~Silagadze,
Int.\ J.\ Mod.\ Phys.\ D {\bf 14}, 143 (2005) [astro-ph/0404515].


\bibitem{book5}
B.~H.~Bransden and C.~J.~Joachain, ``Physics of Atoms and Molecules'',
{\it Prentice Hall, 2nd Edition (2003)}.

\bibitem{paolo1}
P.~Ciarcelluti and R.~Foot,
Phys.\ Lett.\ B {\bf 690}, 462 (2010) [arXiv:1003.0880].

\bibitem{sal3}
F.~Shankar, A.~Lapi, P.~Salucci, G.~De Zotti and L.~Danese,
Astrophys.\ J.\  {\bf 643}, 14 (2006)
[astro-ph/0601577].


\bibitem{sn}
W.~Li {\it et al.}, 
Mon.\ Not.\ Roy.\ Astron.\ Soc.\  {\bf 412}, 1473 (2011) [arXiv:1006.4613].


\bibitem{bp}
A.~Burkert,
IAU Symp.\  {\bf 171}, 175 (1996) [Astrophys.\ J.\  {\bf 447}, L25 (1995)] [astro-ph/9504041];
P.~Salucci and A.~Burkert,
Astrophys.\ J.\  {\bf 537}, L9 (2000) [astro-ph/0004397].

\bibitem{core}
P.~Salucci and  M.~De Laurentis,
arXiv:1302.2268 and references there-in.

\bibitem{lapi}
P.~Salucci {\it et al.}, 
Mon.\ Not.\ Roy.\ Astron.\ Soc.\  {\bf 378}, 41 (2007) [astro-ph/0703115].

\bibitem{PSS}
M.~Persic, P.~Salucci and F.~Stel,
Mon.\ Not.\ Roy.\ Astron.\ Soc.\  {\bf 281}, 27 (1996) [astro-ph/9506004].


\bibitem{foot13}
R.~Foot,
arXiv:1303.1727. 


\bibitem{dsds}
F.~Donato and P.~Salucci,
Mon.\ Not.\ Roy.\ Astron.\ Soc.\  {\bf 353}, L17 (2004)
[astro-ph/0403206].


\bibitem{blok}
W.~J.~G.~de Blok, F.~Walter, E.~Brinks, C.~Trachternach, S-H.~Oh and R.~C.~Kennicutt, Jr.,
Astron.\ J.\  {\bf 136}, 2648 (2008)
[arXiv:0810.2100].

\bibitem{ngc3198}
G.~Gentile {\it et al.},
arXiv:1304.4232 .

\bibitem{tf}
R.~B.~Tully and J.~R.~Fisher,
Astron.\ Astrophys.\  {\bf 54}, 661 (1977).

\bibitem{bell}
E. ~F. ~Bell and R.~S.~de Jong,
Astrophys.\ J.\  {\bf 550}, 212 (2001)
[astro-ph/0011493].

\bibitem{t1}
M. ~A. ~W.~Verheijen,
Astrophys.\ J.\  {\bf 563}, 694 (2001) [astro-ph/0108225].

\bibitem{mnrs}
M. J. Meyer, M. A. Zwaan, R. L. Webster, S. Schneider and L. Staveley-Smith,
Mon.\ Not.\ Roy.\ Astron.\ Soc.\  {\bf 391}, 1712 (2008). 




\bibitem{footrec}
R. ~Foot,
arXiv:1305.4316; arXiv:1407.4213 
and references there-in.



\end{thebibliography}
\end{document}